\documentclass[twocolumn,secnumarabic,amssymb, amsmath, nofootinbib,tightenlines,
nobibnotes, aps, prl,showpacs]{revtex4}
\usepackage{graphicx}
\begin{document}

\title{Microwave Spectroscopy of Cold Rubidium Atoms}


\author{V.M. Entin and I.I. Ryabtsev}

\address{Institute of Semiconductor Physics, Siberian Division, Russian Academy of Sciences, Novosibirsk, 630090 Russia}

\date{June 22, 2004}

\begin{abstract}
The effect of microwave radiation on the resonance fluorescence of a cloud of cold $^{85}Rb$ atoms in a
magnetooptical trap is studied. The radiation frequency was tuned near the hyperfine splitting frequency of
rubidium atoms in the 5S ground state. The microwave field induced magnetic dipole transitions between the
magnetic sublevels of the 5S(F=2) and 5S(F=3) states, resulting in a change in the fluorescence signal. The
resonance fluorescence spectra were recorded by tuning the microwave radiation frequency. The observed spectra
were found to be substantially dependent on the transition under study and the frequency of a repump laser used in
the cooling scheme.

\end{abstract}

\maketitle

The method of optical­radio­frequency double resonance provides the basis for atomic frequency
standards~\cite{bib:01}. Microwave radiation, resonant with transitions between the hyperfine levels of alkali
metal atoms (for example, Rb and Cs), induces magnetic dipole transitions, resulting in turn in a change in a
resonance fluorescence or absorption signal at optical transitions from the ground state. In the first
experiments, the effect of microwave fields on absorption and polarization of light from resonance lamps has been
studied~\cite{bib:02}.

Before the advent of magnetooptical traps for laser cooling and capture of atoms~\cite{bib:03}, thermal atomic
gases with a large Doppler broadening have been mainly studied. The first microwave spectroscopic experiments with
cooled atoms have been performed in a magnetic trap with Na atoms~\cite{bib:1}. The fluorescence spectrum of Na
atoms captured in a strong magnetic field ($\sim $2300 G) was studied in the presence of a probe laser field and
probing microwave radiation. However, a strong magnetic field caused the broadening of resonances up to 200 MHz,
which substantially exceeds the natural width of optical transitions.

Later \cite{bib:2}, a microwave field was used to excite transitions between the hyperfine levels of the ground
state of cesium atoms cooled in the optical molasses at the point of intersection of three standing light waves.
The fluorescence signal from atoms, which have been preliminary cooled in the molasses, was studied after the
shutdown of cooling laser beams. This allowed the observation of optically unperturbed microwave resonances of
width as small as a few tens of hertz. The intensity of microwave radiation in these experiments did not exceed a
few tens of nW/cm$^{2}$.

These studies have been further developed in experiments with so­called "atomic fountains" (see, for example,
\cite{bib:3}). Narrowing of the atomic­standard lines was achieved by the Ramsey fringes method during the
round­trip transit of cooled slow atoms in a microwave resonator~\cite{bib:ramsey}. In addition, a microwave field
was used in some papers instead of a repump laser to produce an optical­microwave magnetooptical
trap~\cite{bib:4}.

The aim of the microwave spectroscopy of cooled atoms is, as a rule, the observation of ultranarrow resonances and
the development of a precision atomic clock based on microwave transitions. However, microwave spectroscopy can be
also used to study processes occurring in a cloud of cold atoms. For example, recently microwave spectroscopic
experiments were performed with atoms loaded from a magnetooptical trap to an optical trap with a large frequency
detuning~\cite{bib:5}. The authors of paper \cite{bib:5} also observed narrow microwave resonances ($\sim $500 Hz
and narrower) by switching off coiling lasers during measurements.

The aim of this work was to study the effect of a microwave field on a resonance fluorescence signal in a standard
magnetooptical trap with Rb atoms and to estimate the possibility of using this effect for diagnostics of a cloud
of cold atoms.

\section*{EXPERIMENTAL SETUP}

The Rb atoms were cooled and captured in a magnetooptical trap using a laser setup containing two 780­nm
external­resonator semiconductor lasers and a frequency­locking system based on saturated absorption in optical
cells with Rb atoms~\cite{bib:EIT,bib:SCT}.

Laser cooling was performed using a standard optical scheme consisting of three pairs of orthogonally polarized
laser beams ($\sigma $+, $\sigma $-) crossing at the center of a quartz cell (Fig. 1), which was evacuated by ion
pumps down to a pressure of $<1 \cdot $10$^{-8}$ Torr. The source of Rb atoms was an ampoule containing a natural
mixture of rubidium isotopes at room temperature.

\begin{figure*}[t]
 \includegraphics[width=12cm]{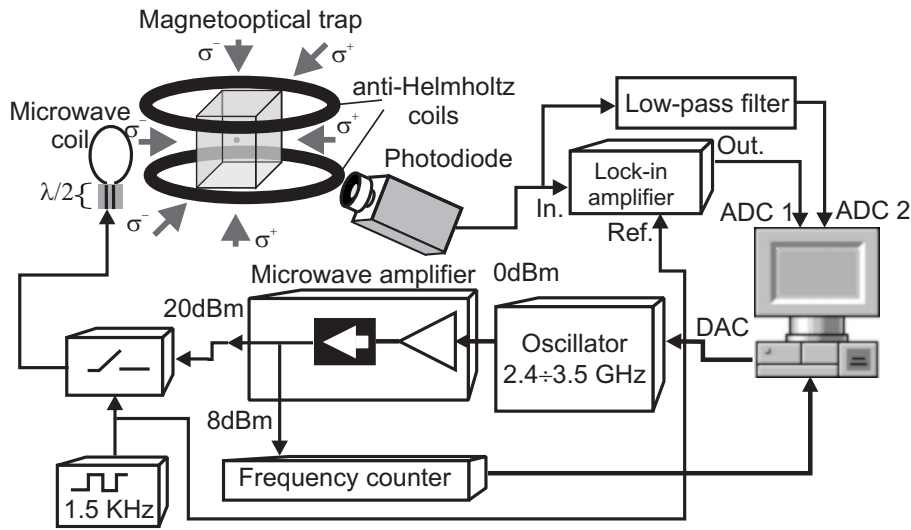}
\caption[figure1]{\label{fig:figure1}Fig. 1. Scheme of the experimental setup.}
\end{figure*}

The magnetic­field gradient in the trap (0$\div$15 G/cm) was produced by a pair of anti­Helmholtz coils. Residual
magnetic fields were compensated with the help of additional Helmholtz coils. The possibility of compensation for
residual magnetic fields and the accuracy of alignment of the laser beams were provided by using two CCD cameras
placed from both sides of the cell.

The resonance fluorescence signal was detected with a calibrated two­element photodiode equipped by a TV objective
and a differential amplifier. The cloud image was projected by the objective on one of the elements of the
photodiode, the second element being used to subtract laser radiation scattered from the cell walls.

The Rb atoms were captured in the trap by locking the cooling laser frequency either to the slope of the saturated
absorption (Fig. 2b) 5S$_{1/2}$(F=3)$ \to $ 5P$_{3/2}$(F=4) transition peak of $^{85}$Rb with the red 1$\div
$3$\Gamma $ detuning from the resonance center (Fig. 2a) ($\Gamma$=6 MHz is the natural width of the D$_{2}$ line
of Rb) or to the slope of the fluorescence resonance of the trap. In the latter case, the so­called
self­stabilized magnetooptical trap was realized~\cite{bib:7}. As a result, the fluorescence signal had a more
stable constant component corresponding to the equilibrium population of the trap.

\begin{figure*}[t]
 \includegraphics[width=14cm]{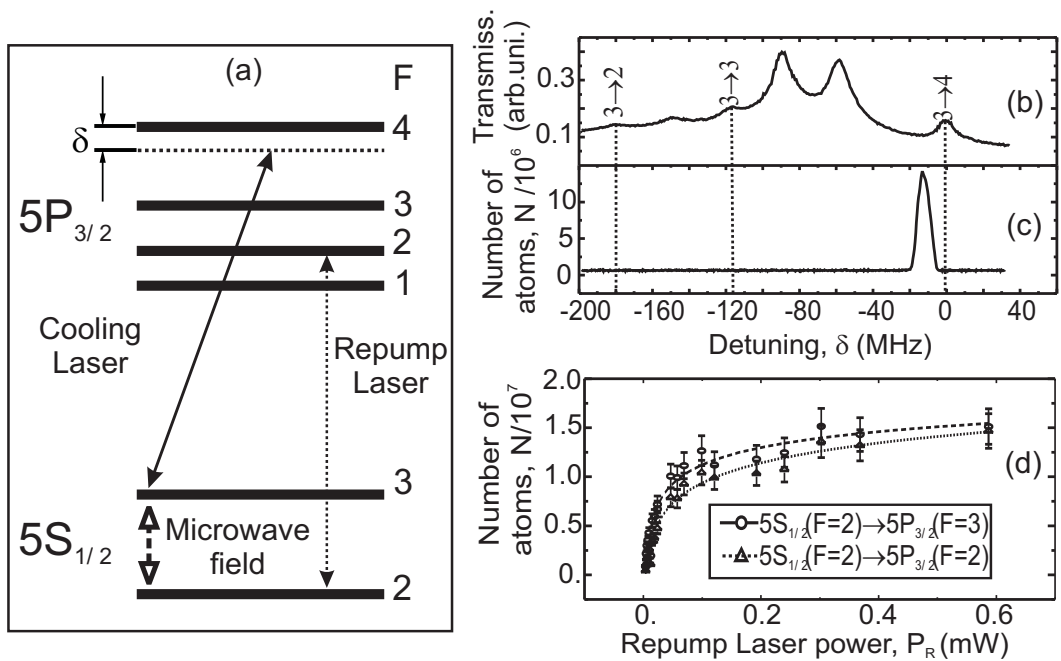}
\caption[figure2]{\label{fig:figure2}Fig. 2. (a) Scheme of transitions in $^{85}$Rb atoms; (b) saturated
absorption spectrum of a reference cell; (c) the number of atoms in the trap as a function of the cooling laser
frequency; (d) dependence of the number of atoms in the trap on the total output power of the repump laser.}
\end{figure*}

The repump laser was tuned to the slope of the saturated absorption 5S$_{1/2}$(F=2)$ \to $ 5P$_{3/2}$(F=2 or 3)
transition peak with the red 0$ \div $3$\Gamma $ detuning (Fig. 2a) and was locked to the resonance using the
Pound-Drever frequency modulation method~\cite{bib:pound1,bib:pound2}.

The output power of lasers was $\sim$6 mW. Laser beams were expanded in front of the cell with the help of
telescopes up to a diameter of 7$ \div $8 mm. The intensity of each of the laser beams incident on the cell was
2.6 mW/cm$^{2}$, while the calculated saturation intensity was 1.65 mW/cm$^{2}$. Figure 2c illustrates the
measured dependence of the number of trapped atoms on the cooling laser frequency, the repump laser frequency
being locked to the resonance at the 5S$_{1/2}$(F=2)$ \to $5P$_{3/2}$(F=2) transition. Figure 2d shows the
dependence of the number of trapped atoms on the total power of the repump laser in the trap when the cooling
laser was detuned by $\delta$=9 MHz. It is seen that the saturation of trap population occurs even at 0.1 mW.

We obtained in the trap a cloud of cold atoms 0.6$ \div $2 mm in diameter, containing $ \le $2$ \times $10$^{7}$
atoms, which corresponds to an atomic density $ \le $2$ \times $10$^{10}$ cm$^{-3}$. The cloud temperature equal
to $ \sim $50$\mu$K was measured by the dynamics of a decrease in the number of atoms after a short (5$ \div $200
ms) switching off of the magnetic field gradient. The trap population decreased by a factor of \textit{e} for the
time t$\sim $30 ms. This means that a greater part of Rb atoms escaped from the region of interaction with laser
radiation for the time 30 ms by flying the distance L=A/2$ \approx $3.5 mm (A is the beam aperture). In the
Maxwell velocity distribution approximation, the expression for the average temperature of atoms can be written in
the form~\cite{bib:1}:

\begin{equation}\label{eq1}
 \langle T\rangle =\frac{M \langle v\rangle ^{2}}{3k_{B}},
\end{equation}

where $\langle $v$\rangle $= L/t is the average velocity of atoms, $k_{B}$ is the Boltzmann constant, and M is the
atom mass; and the temperature estimated from (1) is 50 $\mu $\textit{Ê}.

In microwave spectroscopic experiments, we used a frequency synthesizer based on the first heterodyne of a S4-60
spectrum analyzer with a 20-dBm, 3-GHz ($ \pm $300 MHz) power amplifier (Fig. 1). The width of the heterodyne line
did not exceed 10 kHz. The frequency was continuously tuned with the help of a CAMAC digital­to­analog converter.

A weak change in the resonance fluorescence caused by magnetic dipole transitions in rubidium atoms was
investigated by the lock­in detection technique. For this purpose, the amplitude of microwave radiation was
modulated by a SSW-508 microwave switch (Sirenza). The output microwave radiation was fed from the switch to a
wire coil of diameter 4 cm, which was matched with the help of a strip half­wave transformer and was located at a
distance of 3 cm from the trap center. Control pulses at the frequency 1.5 kHz from a pulse generator were applied
to the microwave switch.

The fluorescence signal from the photodetector was fed to a phase­sensitive amplifier. The output signal of the
generator controlling the switch was used as the reference frequency signal. As a result, the output signal of the
amplifier was proportional to a change in the fluorescence intensity caused by microwave radiation. This signal,
along with a constant fluorescence from the trap, was recorded as a function of the microwave generator frequency.

We have failed in our first experiments to achieve the reproducibility of microwave spectra when the cooling laser
frequency was locked to the saturated absorption resonances. For this reason, we used frequency locking over
fluorescence. Microwave radiation modulated the fluorescence signal at 1.5 kHz. The cut­off frequency of the
cooling laser locking system was selected to be on the order of few hundred of hertz to avoid the suppression of
the useful signal.

To enhance the useful signal, the repump laser power was reduced to 90 $\mu$W for each of the beams in the trap.
In addition, to narrow the microwave resonances in microwave experiments, we used the minimal gradient of the
magnetic field in the trap ($\approx$5G/cm) at which a sufficient number of atoms is still captured by the trap.

\section*{EXPERIMENTAL RESULTS AND DISCUSSION}

Figure 3a shows the dependence of the trap population on the microwave frequency for the case when the repump
laser was detuned to the red with respect to the maximum of the 5S$_{1/2}$(F=2)$ \to $5P$_{3/2}$(F=2) transition.
The microwave spectrum was obtained after averaging over ten measurements. Noise and fluctuations were mainly
caused by the photodetector noise. The position of the frequency 3035.732 MHz of the unperturbed microwave
resonance for the magnetic dipole 5S$_{1/2}$ (F=2)$ \leftrightarrow $ 5S$_{1/2}$(F=3) transition in $^{85}$Rb
atoms is indicated by the dashed straight line. One can see that the fluorescence signal from the trap exhibits a
dip near this frequency. The width of the microwave resonance was almost the same in different experiments and was
$\sim$500 kHz.

\begin{figure*}[t]
 \includegraphics[width=14cm]{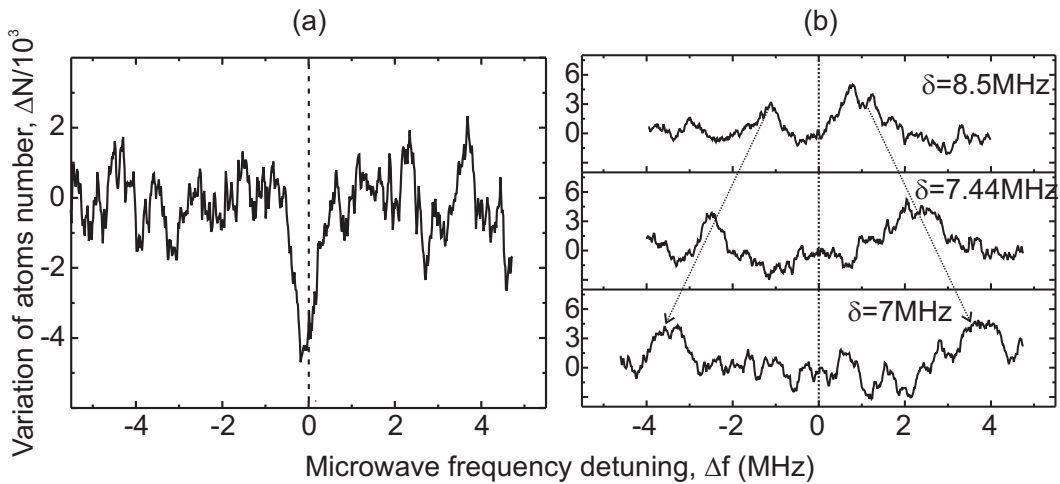}
\caption[figure3]{\label{fig:figure3}Fig. 3. Dependence of the trap population on the microwave-field frequency
for the repump laser tuned to the (a) 5S$_{1/2}$ (F=2)$ \to $5P$_{3/2}$ (F=2) and (b) 5S$_{1/2}$ (F=2)$ \to
$5P$_{3/2}$ (F=3) transitions.}
\end{figure*}

The observed resonance is caused by the transfer of cold $^{85}$Rb atoms from the 5S$_{1/2}$ (F=3) state to the
5S$_{1/2}$ (F=2) state due to magnetic dipole transitions. In the case of exact tuning of the microwave frequency,
the fluorescence intensity of the trap at the cooling 5S$_{1/2}$(F=3)$ \to $5P$_{3/2}$(F=4) transition decreases
because the number of atoms in the 5S$_{1/2}$(F=3) state decreases. The absence of the shift of the microwave
resonance suggests that either the fluorescence signal is detected from the central part of a cloud of cooled
atoms, where the magnetic field is weak and the Zeeman shift of levels is negligible, or magnetic dipole
transitions are excited between the central Zeeman components 5S$_{1/2}$(F=2,$|m_{F}|$=0) and
5S$_{1/2}$(F=3,$|m_{F}|$=0), which are not shifted. The latter assumption is unlikely because the magnetic moments
of cold atoms have no definite orientation with respect to any arbitrary quantization axis in the trap, while the
polarization of microwave radiation has not been especially selected. Therefore, the most probable reason for the
absence of the shift of the microwave resonance is the detection of variations in the fluorescence intensity from
the central part of the cloud. Such a behavior is observed only when the repump laser is tuned to the
5S$_{1/2}$(F=2)$ \to $5P$_{3/2}$(F=2) transition, at which a part of atoms at the trap center can be transferred
to the dark states, which do not interact with laser radiation~\cite{bib:tum}.

The decrease in the fluorescence intensity is most likely explained by the fact that microwave radiation transfers
a part of atoms from the 5S$_{1/2}$(F=3) state to the local dark states, which are produced at the 5S$_{1/2}$(F=2)
level by the repump laser. Because atoms in these states are not excited by the pump laser, they do not also
interact with the radiation of the cooling laser, which results in the reduction of the fluorescence signal.

By analyzing the width of the observed resonance, note that the 5S$_{1/2}$ (F=3) sublevel should experience the
shift, broadening, and splitting under the action of radiation from the cooling laser due to the Autler-Townes
effect~\cite{bib:Autler}. The level splitting and shift are determined by the position of the quasi­energy
levels~\cite{bib:Akulin}:

\begin{equation}\label{eq2}
 \omega_{\pm}=\delta/2\pm \sqrt {\delta ^{2}/4 + \Omega ^{2}/4},
\end{equation}

where $\delta $ is the detuning from the optical resonance and $\Omega $ is the Rabi frequency. Note that the
detuning and Rabi frequency in our experiment (in the case of saturation) are of the same order of magnitude as
$\Gamma$. The magnetic­field gradient determines the variation of detunings on the cloud dimensions, therefore a
microwave resonance should broaden up to a few megahertz. This conclusion is confirmed by recent
paper~\cite{bib:8} in which the Autler-Townes effect was studied on optical transitions in cold Rb atoms and level
shifts and splittings were observed.

However, the resonance width $\approx$400 kHz observed in our experiments is noticeably smaller than the above
value and is virtually independent of the detuning of the cooling laser. This is probably explained by the fact
that microwave radiation has induced transitions to the dark states, which do not interact with radiation. In this
case, the resonance width was determined only by the inhomogeneity of the magnetic field at the trap center, where
the dark states for degenerate levels can appear. This conclusion requires further experiments to study the
features of the dark states produced at the trap center when the repump laser operates at the 5S$_{1/2}$(F=2)$ \to
$5P$_{3/2}$(F=2) transition.

The resonance amplitude in Fig. 3a in different experiments was (3$ \div $9)$ \cdot $10$^{3}$ atoms. Its maximum
value was determined by the strength of the magnetic component of the microwave field. At the same time, it was
found that a rather large variation in the amplitude was caused by weak fluctuations in the detunings of laser
frequencies from optical­transition frequencies. The maximum amplitude was achieved when these detunings were
identical, the central frequency of the resonance being invariable.

Quite a different picture was observed when the repump laser was tuned to the 5S$_{1/2}$(F=2)$ \to
$5P$_{3/2}$(F=3) transition, for which the dark states are absent (Fig. 3b). The spectrum exhibits two peaks ($\mp
\Delta f$), which are symmetrically split with respect to the center of the 5S$_{1/2}$(F=2)$ \leftrightarrow
$5S$_{1/2}$(F=3) transition. It was assumed first that these peaks appear due to the Autler-Townes effect for the
5S$_{1/2}$(F=3) level in the field of the cooling laser because the sifts of resonances depended on the detuning
of the cooling laser at the 5S$_{1/2}$(F=3)$ \to $5P$_{3/2}$(F=4) transition. However, one can see from Eq. (2)
that the shifts of the quasi­energy levels should be different in the case of the red detuning of the laser.
Therefore, the shifts of the resonances in Fig. 3b should be also different, whereas we observe the symmetric
shifts and splitting.

\begin{figure}[t]
 \includegraphics[width=8cm]{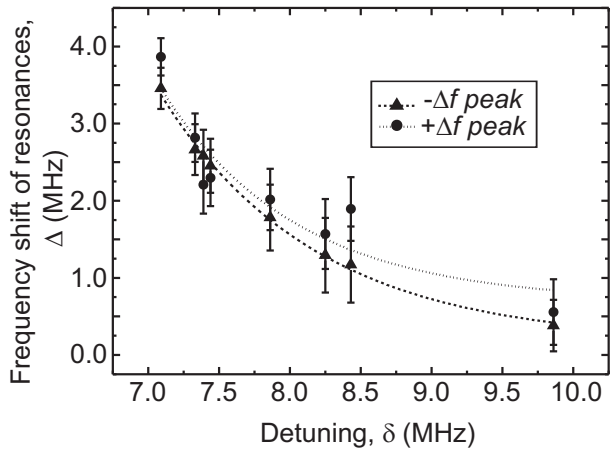}
\caption[figure4]{\label{fig:figure4} Fig. 4. Frequency shift of resonances in Fig. 3b as a function of the
detuning $\delta$ of the cooling laser. The dashed curves are the approximations by the method of least squares. }
\end{figure}

The observed effect can be also interpreted in a different way. Because the frequencies of microwave resonances
are shifted with respect to the central frequency, they are mainly caused by atoms located in the gradient
magnetic field at the periphery of a cold cloud. As the red detuning of the cooling laser from the resonance with
the 5S$_{1/2}$(F=3)$ \to $5P$_{3/2}$(F=4) transition decreases, the temperature of atoms increases, resulting in
an increase in the cloud size. Therefore, the shifts of microwave resonances increase due to the increase in the
Zeeman splitting at the cloud periphery. The increase in the fluorescence signal in the region of resonances in
Fig. 3b compared to its decrease in Fig. 3a is caused by the increase in the efficiency of the repump laser at the
cloud periphery when microwave radiation is tuned to the resonance. The presence of two resonances in Fig. 3b
suggests that magnetic dipole transitions are excited between the extreme Zeeman components of the
5S$_{1/2}$(F=2,$|m_{F}|$=$\pm$2) and 5S$_{1/2}$(F=3,$|m_{F}|$=$\pm$3) levels having the largest resonance shifts
(d$\nu _{Zeeman}$/dB$ \approx $ 2.56 MHz/G). The frequency shifts shown in Fig. 4 correspond to the trap radius
from 1 to 1.8 mm, in accordance with analysis of the TV image of the trap.

Although the main features of the spectra presented above were reproducible in different experiments, note that
the signal noise and fluctuations were rather large. The signal­to­noise ratio was not substantially improved even
in the case of lock­in detection. This is related to a rather long loading time of the trap ($\sim$1 s), resulting
in the suppression of the alternate component of the fluorescence signal. For this reason, the amplitude of
resonances in our experiments did not exceed 0.01 of the total fluorescence intensity level.

At the same time, our experiments showed that the microwave spectroscopy of cold atoms allows one to study
fluorescence signals from different regions of a cloud of cold atoms. The switching to the detection of one or
another region of the cloud is achieved by a proper choice of one of the transitions for the repump laser. For
example, when the J$ \to $J transition is used, a change in fluorescence from the central part of the cloud is
detected, while in the case of the J$ \to $J+1 transitions, fluorescence from the cloud periphery is analyzed.
This is directly related to the presence of the "dark states" for the J$ \to $J transitions and their absence for
the J$ \to $J+1 transitions, as was already mentioned in~\cite{bib:tum}. These features can be used for the
development of new methods for diagnostics of cold atoms.

We thank V.I. Yudin, A.V. Taichenachev, and O.N. Prudnikov for useful discussions. This work was supported by the
Russian Foundation for Basic Research (project no. 02­02­16332) and INTAS (grant no. 2001­155).

\end{document}